\documentclass[bibyear]{aa} 

\usepackage{graphicx}
\usepackage{txfonts}
\usepackage{lipsum}
\usepackage{subcaption}        
\usepackage{lscape}            
\usepackage{placeins}           

\begin{document}

   \title{Constraints on the magnetic field in the hadronic scenario for the origin of the radio halo in the A1795 galaxy cluster}
   \titlerunning{Hadronic scenario for the radio halo in A1795}

\author{P. Marchegiani\inst{1}\corrauth{paolo.marchegiani@inaf.it}       
\and V. Vacca\inst{1}\email{valentina.vacca@inaf.it}        
\and F. Loi\inst{1}\email{francesca.loi@inaf.it}
\and F. Govoni\inst{1}\email{federica.govoni@inaf.it}
\and M. Murgia\inst{1}\email{matteo.murgia@inaf.it}
}

   \institute{INAF-Osservatorio Astronomico di Cagliari, Via della Scienza 5, I-09047 Selargius (CA), Italy}

   \date{Received 25 May 2026 ; Accepted 25 July 2026}

  \abstract 
   {The radio halo in the cool-core galaxy cluster, A1795, has recently been proposed to be of hadronic origin, because it extends on megaparsec scales, where no turbulence is expected to be injected into a relaxed cluster.}
   {In this paper we aim to constrain the properties of nonthermal protons and the cluster magnetic field strength under the hypothesis of a hadronic origin for the radio halo in the light of the available gamma-ray upper limits for this cluster.}
   {We assumed a radial profile and several central magnetic field intensity values, then derived the corresponding properties of the nonthermal protons necessary to reproduce the flux density and surface brightness profile of the radio halo. For the obtained results, we calculated the corresponding gamma-ray emission and compared it with the \textit{Fermi}-LAT upper limit.}
   {We find that, for a magnetic field energy radial profile following that of the thermal gas, central magnetic field intensities $B_0<5\,\mu$G imply a gamma-ray emission at or above the \textit{Fermi}-LAT upper limit. Therefore, larger values of the magnetic field are necessary in the hadronic scenario. For these values the radial profile of nonthermal protons must slightly decrease with radius, though more slowly than that of the thermal gas. For smaller values of the magnetic field, different mechanisms for producing or accelerating the nonthermal electrons are necessary.}
  {}


   \maketitle
\nolinenumbers

\section{Introduction}

Radio halos in galaxy clusters comprise diffuse emissions with shapes and sizes similar to those of thermal X-ray emission. They reveal cluster-scale relativistic electrons and magnetic fields (e.g., Feretti et al. 2012; van Weeren et al. 2019).

Since the lifetime and diffusion length of relativistic electrons are shorter than the cluster's age and size, the electrons must be continuously produced or accelerated on widespread scales (e.g., Brunetti \& Jones 2014 for a review). Proposed mechanisms include hadronic interactions between nonthermal protons and nuclei in the intracluster medium (Dennison 1980; Blasi \& Colafrancesco 1999) and reacceleration induced by cluster turbulence (Tribble 1993; Brunetti et al. 2001). Such turbulence can develop on large scales after a cluster merger or on smaller scales due to other mechanisms such as gas sloshing or active galactic nuclei activity (Vazza et al. 2012). 
Since radio halos are usually observed in disturbed clusters (Buote 2001; Cassano et al. 2013) and the \textit{Fermi}-Large Area Telescope (LAT) did not detect gamma-ray emission even in massive and nearby clusters (Ackermann et al. 2014) as expected in the hadronic scenario (Marchegiani et al. 2007), the turbulence scenario is typically considered to be a viable solution for explaining the origin of radio halos.

However, using the LOw Frequency ARray (LOFAR) and MeerKAT, van Weeren et al. (2026; hereafter W26) recently detected a giant radio halo, with a size of the order of 1 Mpc, in the cool-core galaxy cluster A1795. This cluster appears relaxed, with no evidence of an ongoing or recent merger. 
This case is slightly similar to that of CL1821+643, another cool-core cluster with a detected megaparsec-sized radio halo (Bonafede et al. 2014). However, one of the parameters often used to infer the dynamical state of a cluster -- the centroid shift $w$ (Cassano et al. 2010) -- has a value typical of disturbed clusters when calculated in CL1821+643. This suggests that a possible off-axis or minor merger may be ongoing in that cluster. In A1795, by contrast, both the centroid shift ($w$) and the concentration parameter ($c$) are typical of relaxed clusters (Botteon et al. 2022), suggesting that a similar explanation is not viable. This has also been confirmed by a recent XRISM result (Sarkar et al. 2026), which reveals a very low level of turbulence in this cluster, even compared to other cool-core clusters.
Therefore, W26 hypothesized a possible hadronic origin for the halo in A1795, distinguishing it from other known halos. This would provide an opportunity to study nonthermal cluster components such as nonthermal protons and secondary electrons in their equilibrium state without the modifications potentially introduced by turbulent reacceleration (e.g., Brunetti et al. 2012). 

In the hadronic scenario, degeneracy between the properties of the nonthermal protons and magnetic field can be partially disentangled by using information about the radial properties of the radio halo (Marchegiani et al. 2007). Currently, there is a limited available information about the magnetic field in A1795 (Ge \& Owen 1993; see also the databases of Osinga et al. 2022, 2025); therefore, W26 calculated the expected radio halo surface brightness profile in the hadronic scenario by considering several possibilities for the central magnetic field intensity and the radial profile of both the nonthermal protons and the magnetic field. They find that the best agreement with the observed radio halo profile is obtained for a radially flat distribution of nonthermal protons and a magnetic field energy radial profile following that of the thermal gas, i.e., $B(r)\propto n_{th}^{0.5}(r)$. Their derived best-fit central magnetic field values range between 1 and 5 $\mu$G. While higher values are allowed in their paper, they are disfavored because they would require assuming a radially decreasing profile for the nonthermal protons.

An important prediction of the hadronic scenario is gamma-ray emission, produced via the same hadronic interactions that generate the secondary electrons (Colafrancesco \& Blasi 1998). 
Using this property, gamma-ray upper limits in several clusters, combined with radio data, have been used to constrain the magnetic field intensity, assuming a pure hadronic scenario. This approach yields lower limits of the order of microgauss (e.g., Jeltema \& Profumo 2011). Predictions for gamma-ray emission have been derived for the well-studied Coma galaxy cluster, considering hadronic scenarios that include other effects. These effects may involve the combination of primary and secondary electrons (Zandanel et al. 2014; Adam et al. 2021) or the turbulent reacceleration of secondary electrons (e.g., Brunetti et al. 2012, 2017; Marchegiani et al. 2026), and have even been explored using numerical simulations (e.g., Pinzke et al. 2017). For the specific case of A1795, predictions of gamma-ray emission were derived under two different assumptions: (1) cosmic-ray heating balances thermal gas cooling (Colafrancesco \& Marchegiani 2008), or (2) a mix of primary and secondary electrons (Adam et al. 2020). In both cases these predictions were made before the giant radio halo was detected by W26 and therefore did not incorporate comparisons with the giant halo's radio data.
In this paper we explore the constraints that can be obtained by augmenting the pure hadronic scenario for the giant radio halo in A1795, studied by W26, with the information derived from the \textit{Fermi}-LAT upper limit for this cluster (Ackermann et al. 2014). Our goal is to obtain additional constraints on the properties of nonthermal protons and the magnetic field.

Below, in Sect. 2, we describe the assumptions and methods used in our calculations. We present our results in Sect. 3 and summarize our findings and conclusions in Sect. 4. For direct comparison with the results of W26, we adopt the same cosmological model as in their paper, i.e., a $\Lambda$-cold dark matter model with $\Omega_m = 0.3$, $\Omega_{\Lambda} = 0.7$, and $H_0 =70$ km s$^{-1}$ Mpc$^{-1}$. For this cosmological model, the luminosity distance of the A1795 cluster, located at $z=0.0625$, is $D_L=280$ Mpc, and 1 arcmin corresponds to 72.2 kpc.

\section{Methods}

To model the hadronic scenario, we followed the approach described in Marchegiani et al. (2007). This method is based on the calculated production rate of relativistic electrons, positrons (hereafter referred to simply as electrons), and gamma rays from interactions between nonthermal protons and thermal nuclei, as presented by Moskalenko and Strong (1998) and Furlanetto and Loeb (2002). These calculations, in turn, are based on the model for charged and neutral pion production in hadronic interactions developed by Dermer (1986a,b).

We described the nonthermal protons energy spectrum as a power law and their radial profile as proportional to the thermal profile raised to a power $\xi$, with their normalization $N_{p,0}$ left as a free parameter:
\begin{equation}
N_p(\gamma,r)=N_{p,0}\gamma^{-s_p}[n_{th}(r)/n_{th,0}]^{\xi}.
\label{eq.density_scaling}
\end{equation}
Similarly, we described the magnetic field as proportional to the thermal profile raised to a power $\eta$:
\begin{equation}
B(r)=B_0 [n_{th}(r)/n_{th,0}]^{\eta}.
\label{eq.magnetic_scaling}
\end{equation}
For the thermal profile, we used the parameterization provided by Vikhlinin et al. (2006) derived from Chandra data. We assumed that the radial distributions of nonthermal protons and magnetic field extend up to a radius of 600 kpc, as shown for the radio halo surface brightness in Fig. 8 in W26.
The spectral index of nonthermal protons was set at $s_p=2.2$ to reproduce the radio halo spectral index $\alpha\sim1.1$ derived by W26 between 144 and 1279 MHz.

With these assumptions, we calculated the source spectrum of secondary electrons $Q_e(\gamma,r)$ and from it the spectrum and the radial distribution of the nonthermal electrons, $N_e(\gamma,r)$, resulting from the equilibrium between continuous production via hadronic interactions and energy losses:
\begin{equation}
N_e(\gamma,r)=\frac{1}{b(\gamma,r)}\int^{\infty}_{\gamma}Q_e(\gamma',r) d \gamma'
\end{equation}
(e.g., Sarazin 1999), where $b(\gamma,r)$ is the energy loss term. For electrons with energies $\gamma>10^3$, this term is dominated by synchrotron interactions with the magnetic field and inverse Compton scattering with cosmic microwave background photons.

From the electron equilibrium spectrum, we calculated the flux density radio spectrum and the surface brightness profiles at 144 and 1279 MHz, resulting from the synchrotron interaction between nonthermal electrons and the magnetic field. 
In this calculation, we set the radial shape of the magnetic field to $\eta=0.5$, corresponding to a magnetic field energy that scales with the thermal profile. This choice was motivated by rotation measure (RM) analysis of the Coma cluster from  (Bonafede et al. 2010), studies of the radial profile of diffuse emission in A2744 (Murgia et al. 2009), and the fact that this scaling provides the best fit for the radio halo surface brightness profile in A1795 (W26).
In Appendix A, we also explore cases with different values of $\eta$.
We varied the central magnetic field $B_0$ across the $1-10$ $\mu$G range and, for each value of $B_0$, considered several values of the parameter $\xi$, searching for the best fit to reproduce the radio halo's surface brightness profile. For comparison, we used the data at the two frequencies plotted in Fig. 8 of W26. We determined the value of $N_{p,0}$ by requiring that the model reproduce the radio halo's total flux density values at the two frequencies reported in Table 3 of W26. Parameters were chosen based on visual inspection. A more quantitative comparison based on statistical estimators would require a wider inspection of all possible parameter variations and, hence, much longer calculation times. This could be the subject of a future paper.

Finally, for the parameter values obtained through this method, we calculated the gamma-ray emissivity $Q_g(E,r)$ and the corresponding flux in the energy range $E>500$ GeV emitted by the cluster. We then compared our results with the \textit{Fermi}-LAT upper limit reported by Ackermann et al. (2014): $F_{UL}(>500\mbox{ MeV})= 3.01\times10^{-10}$ cm$^{-2}$ s$^{-1}$.

\section{Results}

Our results are summarized in Table \ref{Table1}. For each considered value of $B_0$, we report the best-fit values of $\xi$ and $N_{p,0}$ that reproduce both the observed surface brightness profile and the flux density of the radio halo, together with the corresponding gamma-ray flux.
From these results, we see that the predicted gamma-ray emission equals the \textit{Fermi}-LAT upper limit for a given central magnetic field value between 4 and 5 $\mu$G. Consequently, central magnetic field values of the order of or greater than 5 $\mu$G are necessary to reproduce the properties of the radio halo in the hadronic scenario without violating the gamma-ray upper limit.

\begin{table}[t]
\caption{Parameters and results.}
\label{Table1}
\centering
\begin{tabular}{cccc}
\hline\hline
$B_0$ & $\xi$ & $N_{p,0}$ & $F(>500\mbox{ MeV})$ \\
($\mu$G) & & (cm$^{-3}$) & (cm$^{-2}$ s$^{-1}$) \\
\hline
1 & 0.2 & $5.5\times10^{-8}$ & $5.9\times10^{-9}$ \\
3 & 0.2 & $6.2\times10^{-9}$ & $6.7\times10^{-10}$ \\
4 & 0.25 & $4.0\times10^{-9}$ & $3.7\times10^{-10}$ \\
5 & 0.3 & $3.1\times10^{-9}$ & $2.4\times10^{-10}$ \\
7 & 0.4 & $2.3\times10^{-9}$ & $1.3\times10^{-10}$ \\
10 & 0.5 & $1.8\times10^{-9}$ & $7.4\times10^{-11}$ \\
\hline
\end{tabular}
\tablefoot{For $\eta=0.5$, we list the parameter values that best reproduce the radio halo's flux density, surface brightness profile, and corresponding gamma-ray emission.}
\end{table}

\begin{figure}[t]
    \centering
    \includegraphics[width=\hsize]{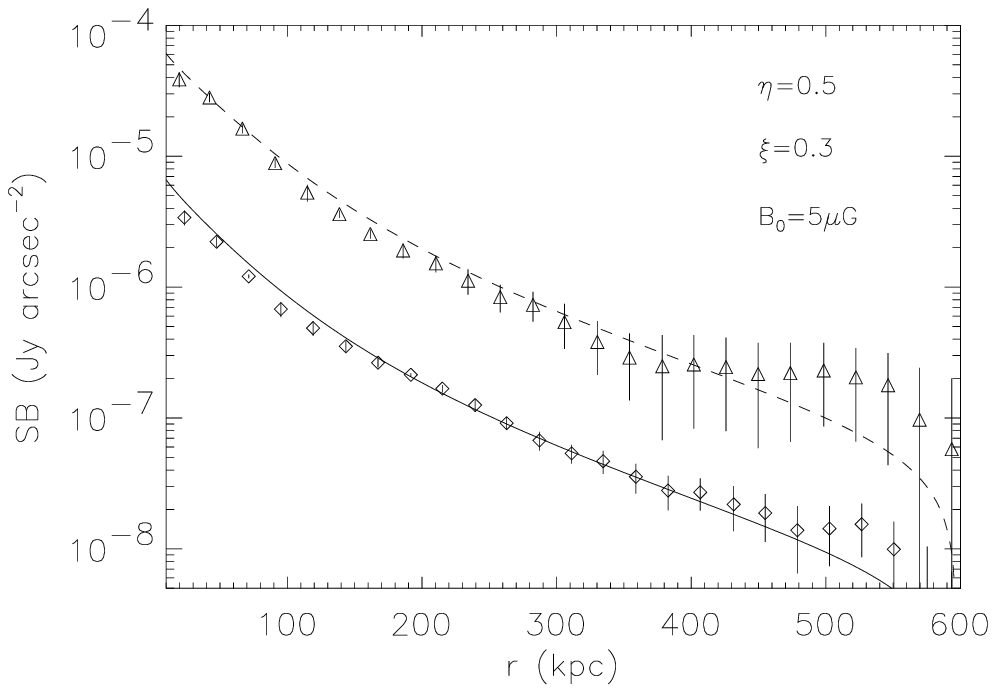}
       \caption{Surface brightness radial profiles for 
$B_0=5$ $\mu$G, $\eta=0.5$, $\xi=0.3$, and $N_{p,0}=3.1\times10^{-9}$ cm$^{-3}$
at 1279 MHz (solid line) and 144 MHz (dashed line), compared with the data in Fig. 8 of W26 from MeerKAT (diamonds) and LOFAR (triangles).}
          \label{fig1}
    \end{figure}

We also observe that for central magnetic field values below 3 $\mu$G, the best agreement with the radial profile of the radio halo surface brightness is achieved for $\xi=0.2$, corresponding to a radial distribution of nonthermal protons that decreases slowly toward the cluster periphery. For higher values of $B_0$, higher values of $\xi$ up to 0.5 are necessary. This implies a steeper radial distribution of nonthermal protons, which, in any case, is flatter than that of the thermal one. This aligns with the results of W26 and arises because for $B>3$ $\mu$G synchrotron energy losses dominate over inverse Compton scattering losses. This suppression of the equilibrium nonthermal electron density at the cluster center thus requires a steeper distribution of nonthermal protons to compensate.  

In Fig.\ref{fig1} we show an example, for the case with $B_0=5\,\mu$G, of the resulting 
surface brightness radial profiles compared with the data presented in W26. We show only this case because the results for all other cases are visually very similar, differing only in the parameter values reported in Table \ref{Table1}.

We also note that, while these models are compatible with most of the error bars, they predict a surface brightness profile that does not fit the observed data very well at large distances from the center. This may suggest that the nonthermal protons extend beyond the previously assumed 600 kpc radius, resulting in a less steep decline of the surface brightness profile at large radii. Alternatively, it may suggest the presence of an additional electron component of different origin emerging at large distances from the cluster center (see, e.g., Zandanel et al. 2014). 

\begin{figure}[t]
    \centering
    \includegraphics[width=\hsize]{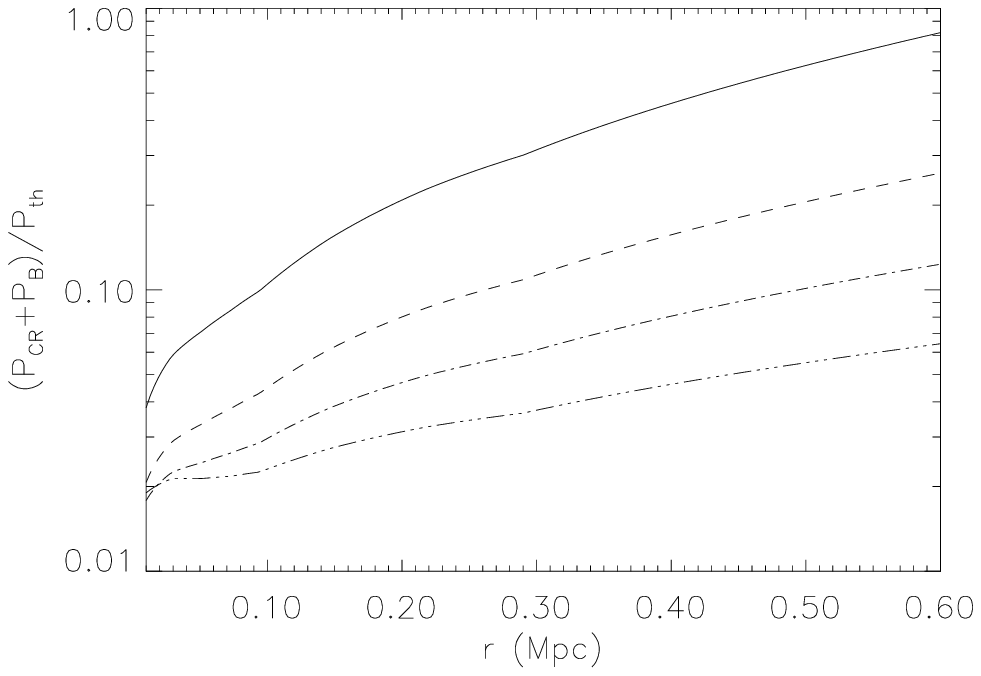}
       \caption{Pressure ratio $(P_{CR}+P_B)/P_{th}$ as a function of radius for the models shown in Table \ref{Table1}. The curves correspond to cases with $B_0=3$$\mu$G (solid), 5$\mu$G (dashed), 7$\mu$G (dot-dashed), and 10 $\mu$G (three dots-dashed).}
          \label{fig2}
    \end{figure}

In Fig.\ref{fig2} we also show, for cases with $B_0\geq 3$ $\mu$G, the radial trend of the ratio between the nonthermal pressure (i.e., the sum of the cosmic-ray pressure $P_{CR}$ and the magnetic field pressure $P_B$), and thermal pressure, $P_{th}$. This plot shows that, in all cases, the ratio increases toward the cluster periphery, with higher values obtained for smaller magnetic fields. In particular, for $B_0=3$ $\mu$G the pressure ratio at the periphery reaches values of the order of $80\%$, while for higher magnetic fields it remains under $30\%$. We do not show the case with $B_0=1$ $\mu$G, where the pressure ratio at the periphery reaches values of the order of $700\%$, as it would lie outside the area shown in this plot. Therefore, values of the central magnetic field greater than 5 $\mu$G are also preferred by this pressure ratio analysis, as high pressure ratios would be difficult to explain, particularly in a cluster lacking evident sources of cosmic-ray acceleration at high distances from the center, such as large shock fronts. 

\section{Discussion and conclusions}

These results confirm that, in the hadronic scenario for the origin of the radio halo in A1795, if we assume a magnetic field with an energy profile following that of the thermal gas, the nonthermal protons must have a radially flatter distribution than the thermal gas. Slightly steeper profiles are required for higher values of the central magnetic field, consistently with the findings of W26.

In this work, we find that, to avoid violating the \textit{Fermi}-LAT upper limit for this cluster, values of the central magnetic field of the order of $1-4$ $\mu$G can be excluded, requiring values of 5 $\mu$G or higher. 
This result holds when assuming $\eta=0.5$, which, as previously discussed, is the preferred case in W26 and in other studies of diffuse emissions and RMs in clusters. In principle, other values of this parameter are possible, as discussed in Appendix A.

We note that the framework adopted in this paper is similar, but not exactly identical to that of W26. Specifically, in that paper the energy density of nonthermal protons, $W_{nth}$, is parameterized by scaling it relative to the thermal energy density, $W_{th}$, through an exponent, $a$, defined as $W_{nth}\propto W_{th}^a$. In our framework the exponent, $\xi$, refers to the scaling of numerical densities (see Eq.\ref{eq.density_scaling}). Another difference lies in the parameterization of the thermal gas profile: we derived ours from Vikhlinin et al. (2006), whereas W26 used the parameterization of Andrade-Santos et al. (2017), in which the values of the profile parameters are not explicitly given. Thus, a direct comparison is not possible. However, both Vikhlinin et al. and Andrade-Santos et al. used Chandra data, the same fitting procedure, and identical functional forms, so we do not expect the differences to be significant. These different assumptions may explain some of the numerical differences between the results of this paper and W26, such as the discrepancy in the derived best-fit shape of the nonthermal protons profile. For $\eta=0.5$ and $B_0=5$ $\mu$G, we find $\xi=0.3$, whereas W26 report $a=0$. Nevertheless, the resulting radio surface brightness profiles are quite similar in the two papers, except at radii close to 600 kpc. Here, the difference may arise from differences in the assumed maximum radius to which nonthermal protons extend (fixed at 600 kpc in this paper, due to the lack of information at larger radii). The similarity of the two profiles suggests that, while some differences in the exact numerical values of the central magnetic field might emerge if had adopted exactly the same assumptions as W26, such differences would be small. 

We also note that central magnetic field values higher than 10 $\mu$G have been found in cool-core clusters such as Hydra (Laing et al. 2008) and A2199 (Vacca et al. 2012); thus, the constraint obtained in this paper aligns with that found in other clusters with similar properties (see Vacca et al. 2026 for a discussion about the properties of the magnetic field derived in clusters with different dynamical states). 
Moreover, the central magnetic field intensity statistically correlates with the central density of the thermal gas, as shown by Govoni et al. (2017), Loi et al. (2026), and Vacca et al. (2026). These studies predict a mean central magnetic field ($B_0$) of about 10--15 $\mu$G for a central thermal gas density $n_0\sim4\times10^{-2}$ cm$^{-3}$ (i.e., the central density value found for A1795 by Vikhlinin et al. 2006). This prediction aligns with the constraints found in this paper.

Therefore, independent measurements of the magnetic field, such as those derived from Faraday rotation analyses, may help verify the viability of the hadronic scenario for explaining the origin of the radio halo in this cluster. A central magnetic field value of the order of 4 $\mu$G or less would allow us to exclude the pure hadronic scenario and imply that other mechanisms may produce the radio halo in this cluster. In this respect, we note that there are several polarized sources in the A1795 region (six sources within a projected distance of 1.2 Mpc from the cluster center, according to the databases of Osinga et al. 2022, 2025), making an RM study feasible, in principle. In the past, a study of the magnetic field properties using Faraday RM in this cluster was conducted by Ge and Owen (1993), who estimated a magnetic field strength greater than 20 $\mu$G -- albeit with large uncertainties -- using VLA observations of the central galaxy. This estimate aligns with the constraints obtained in this paper. In the future, highly sensitive instruments such as MeerKAT+ or SKA-mid could increase the number of background polarized sources detected in that region, because this number is approximately inversely proportional to the sensitivity (Loi et al. 2019). Therefore, such observations may enable a solid determination of the magnetic field properties.

Another possibility to test the hadronic scenario for the formation of the radio halo in this cluster is to observe it at higher frequencies, for example, in the C band. In fact, the hadronic scenario predicts a spectrum that maintains a power-law shape up to high frequencies,  
unlike turbulent reacceleration models, which instead expect a steepening (e.g., Brunetti et al. 2001). Therefore, confirming a power-law spectral shape at higher frequencies would provide important information to test the hadronic scenario.

\begin{acknowledgements}
We thank the referee for useful comments and suggestions.      
\end{acknowledgements}

\begin{appendix}

\onecolumn
\nolinenumbers

\section{Different radial profiles of the magnetic field}

\begin{figure*}[ht]
    \centering
  \hbox{  \includegraphics[width=0.25\textwidth]{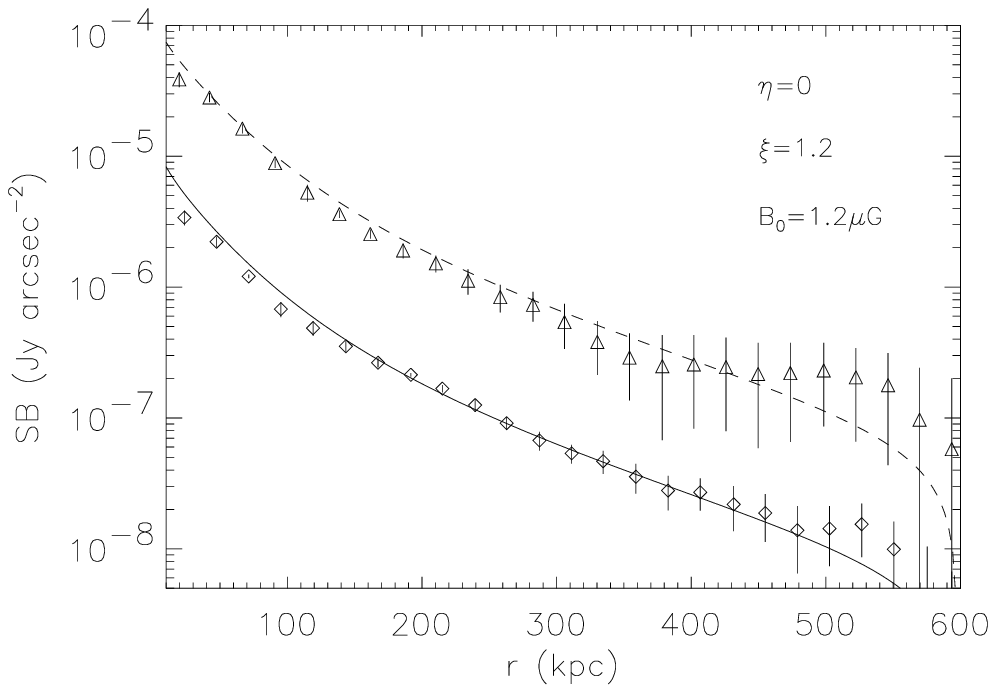}
    \includegraphics[width=0.25\textwidth]{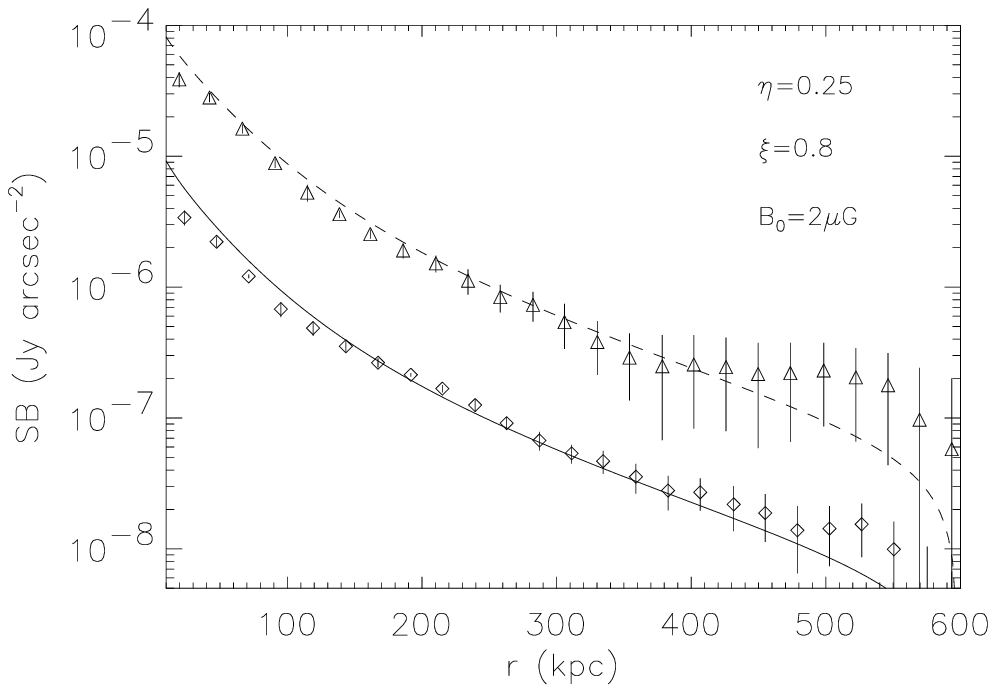} 
      \includegraphics[width=0.25\textwidth]{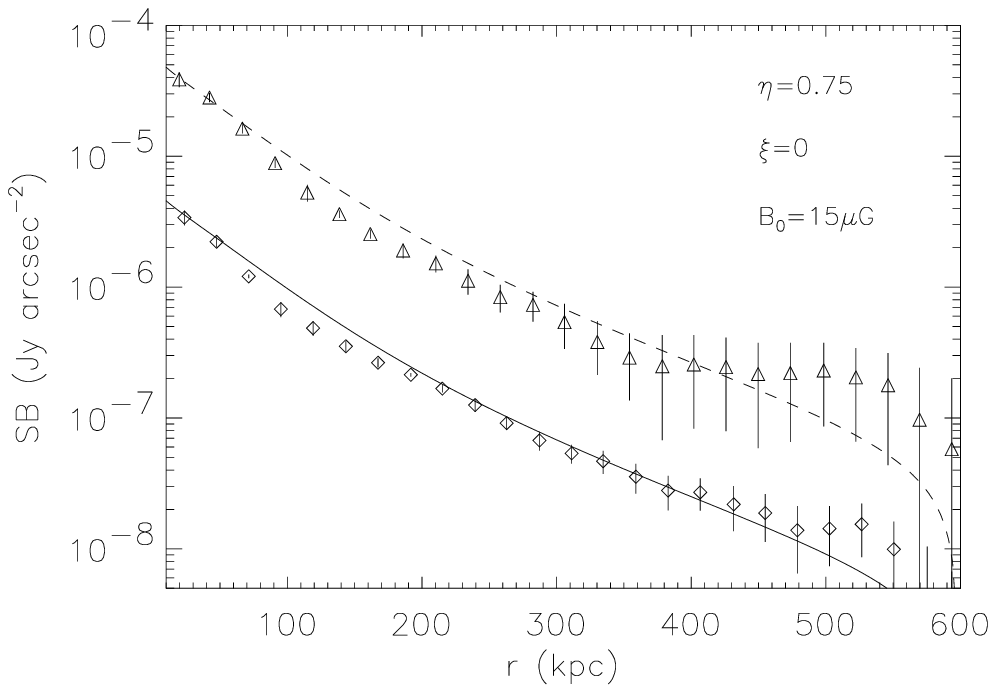}
    \includegraphics[width=0.25\textwidth]{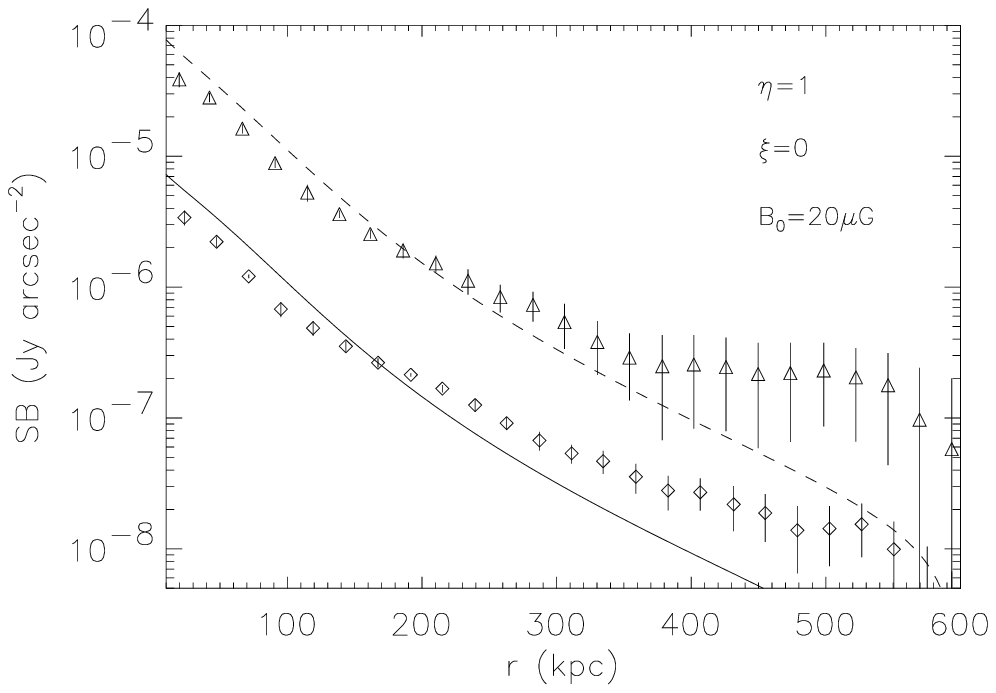} }
       \caption{Surface brightness profiles of the radio halo at 144 MHz (dashed line) and 1279 MHz (solid line), compared with LOFAR and MeerKAT data respectively, for different values of $\eta$ and the corresponding values of $\xi$ and $B_0$ as given in the labels inside the plots.}
          \label{fig.1app}
    \end{figure*}

In this Appendix we consider some cases with different values of the parameter $\eta$ describing the scaling of the radial profile of the magnetic field with respect to the thermal one (see eq.\ref{eq.magnetic_scaling}). While $\eta=0.5$ is our benchmark case, we consider cases with $\eta$ between 0 and 1, i.e., $\eta=0$, 0.25, 0.75, and 1. For each value of $\eta$, we consider several values of $\xi$ and $B_0$ by searching for the combination of parameters that better reproduces the surface brightness profile of the radio halo at 144 and 1279 MHz measured by W26, and adjusting the value of $N_{p,0}$ in order to reproduce the observed flux density at the two frequencies. Results are shown in Fig.\ref{fig.1app}.

For the case $\eta=0$ we found that the halo surface brightness profile can be reproduced with $\xi=1.2$, i.e., a profile of nonthermal protons steeper than the thermal one, while for $\eta=0.25$ we found $\xi=0.8$. In these cases the values of the central magnetic field that provide a gamma-ray flux of the order of the \textit{Fermi}-LAT upper limit are $B_0\sim1.2$ and 2 $\mu$G respectively; therefore the energy losses are dominated by inverse Compton scattering and the value of $B_0$ does not affect the shape of the surface brightness profile.

For $\eta=0.75$ we found that for $\xi=0$ (i.e., a constant profile of nonthermal protons) the halo surface brightness profile can be reproduced for a central magnetic field of $B_0\sim15$ $\mu$G; unlike the previous cases, in this case the energy losses are dominated by synchrotron and the value of $B_0$ determines the shape of the surface brightness profile. In this case the lower limit on the central magnetic field determined by the gamma upper limit is $B_0\sim10$ $\mu$G.

For $\eta=1$ we did not find a value of $\xi\geq0$ providing a good fitting to the surface brightness profile for any value of the central magnetic field. Since $\xi<0$ would mean a radial profile of the nonthermal protons increasing toward the cluster periphery, we did not consider this possibility. For $\xi=0$ we found that a gamma-ray flux of the order of the \textit{Fermi}-LAT upper limit is produced for a central magnetic field of the order of 20 $\mu$G. 

These results show that the lower limit to the magnetic field derived in this paper is depending on the value of $\eta$. It is therefore important to perform a detailed characterization of the properties in total intensity and polarization of the radio halo and radio galaxies inside the cluster in order to simultaneously constrain the magnetic field strength at the center of the cluster and its radial decline (as done for example in Govoni et al. 2006 and Vacca et al. 2010). 
In this respect, we note that, as described in Sect.2, the case $\eta=0.5$ is resulting as favorite in several studies of RM and diffuse emission in clusters, and also in the W26 analysis of this cluster. However, other RM studies, stacking over samples of clusters without distinguishing their dynamical states (Osinga et al. 2025), seem to prefer values $\eta<0.5$, whereas studies in individual clusters point toward values close to $\eta=1$ (e.g., Vacca et al. 2012, Govoni et al. 2017). Therefore, while in this paper we consider the case with $\eta=0.5$ as a motivated benchmark case, it should be taken into account that more precise constraints on the magnetic field, as derivable from detailed RM studies, would be important to better constrain the results we have presented.
    
\begin{figure}[hb]
    \centering
    \includegraphics[scale=0.4]{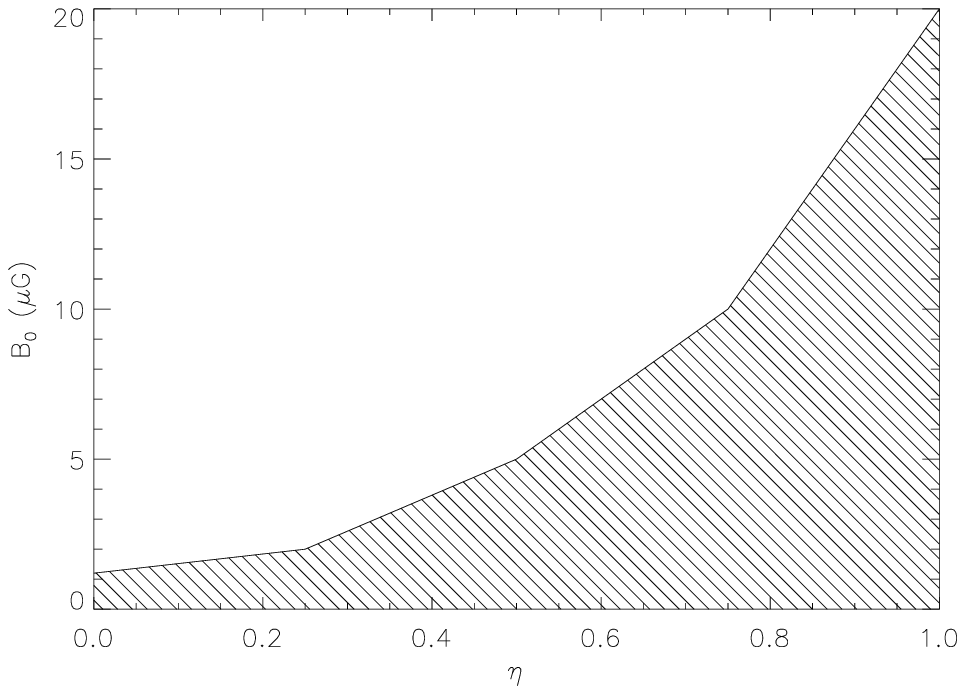}
       \caption{Constraints on the central value of the magnetic field for different values of $\eta$ obtained by requiring that the gamma-ray emission does not violate the \textit{Fermi}-LAT upper limit; the shaded area represents the forbidden region.}
          \label{fig.2app}
    \end{figure}

In order to better visualize the constraints obtained in this paper, we summarize them in Fig.\ref{fig.2app}, where for different values of $\eta$ we show the corresponding values of $B_0$ providing a gamma-ray flux of the order of the \textit{Fermi}-LAT upper limit. In this plot, the shaded area indicates the values of $B_0$ providing a gamma-ray emission in excess.

\end{appendix}


\begin{thebibliography}{}

\bibitem[2014]{Ackermann2014}
Ackermann, M., Ajello, M., Albert, A., et al., 2014, ApJ, 787, 18

\bibitem[2020]{Adam2020} 
Adam, R., Goksu, H., Leing{\"a}rtner-Goth, A., et al., 2020, A\&A, 644, A70 

\bibitem[2021]{Adam2021} 
Adam, R., Goksu, H., Brown, S., et al., 2021, A\&A, 648, A60

\bibitem[2017]{Andrade2017} 
Andrade-Santos, F., Jones, C., Forman, W.~R., et al., 2017, ApJ, 843, 76

\bibitem[1999]{Blasi1999}
Blasi, P., \& Colafrancesco, S., 1999, APh, 12, 169

\bibitem[2010]{Bonafede2010}
Bonafede, A., Feretti, L., Murgia, M., et al., 2010, A\&A, 513, A30

\bibitem[2014]{Bonafede2014} 
Bonafede, A., Intema, H.~T., Bruggen, M., et al., 2014, MNRAS, 444, L44

\bibitem[2022]{Botteon2022}
Botteon, A., Shimwell, T.W., Cassano, R., et al., 2022, A\&A, 660, A78

\bibitem[2014]{Brunetti2014}
Brunetti, G., \& Jones, T.W., 2014, International Journal of Modern Physics D, 23, 1430007

\bibitem[2001]{Brunetti2001} 
Brunetti, G., Setti, G., Feretti, L., \& Giovannini, G., 2001, MNRAS, 320, 365

\bibitem[2012]{Brunetti2012} 
Brunetti, G., Blasi, P., Reimer, O., et al., 2012, MNRAS, 426, 956

\bibitem[2017]{Brunetti2017}
Brunetti, G., Zimmer, S., \& Zandanel, F., 2017, MNRAS, 472, 1506

\bibitem[2001]{Buote2001} 
Buote, D.~A., 2001, ApJL, 553, L15

\bibitem[2010]{Cassano2010}
Cassano, R., Ettori, S., Giacintucci, S., et al., 2010, ApJ, 721, L82

\bibitem[2013]{Cassano2013}
Cassano, R., Ettori, S., Brunetti, G., et al., 2013, ApJ, 777, 141

\bibitem[1998]{Colafrancesco1998} 
Colafrancesco, S., \& Blasi, P., 1998, APh, 9, 227

\bibitem[2008]{Colafrancesco2008} 
Colafrancesco, S., \& Marchegiani, P., 2008, A\&A, 484, 51

\bibitem[1980]{Dennison1980} 
Dennison, B., 1980, ApJL, 239, L93

\bibitem[1986]{Dermer1986a} 
Dermer, C.D., 1986a, ApJ, 307, 47

\bibitem[1986]{Dermer1986b} 
Dermer, C.D., 1986b, A\&A, 157, 223

\bibitem[2012]{Feretti2012}
Feretti, L., Giovannini, G., Govoni, F., \& Murgia, M., 2012, A\&ARv, 20, 54

\bibitem[2002]{Furlanetto2002} 
Furlanetto, S.R., \& Loeb, A., 2002, ApJ, 572, 796

\bibitem[1993]{Ge1993}
Ge, J. P., \& Owen, F. N., 1993, AJ, 105, 778

\bibitem[2006]{Govoni2006} 
Govoni, F., Murgia, M., Feretti, L., et al., 2006, A\&A, 460, 425 

\bibitem[2017]{Govoni2017}
Govoni, F., Murgia, M., Vacca, V., et al. 2017, A\&A, 603, A122

\bibitem[2011]{Jeltema2011} 
Jeltema, T.~E., \& Profumo, S., 2011, ApJ, 728, 53

\bibitem[2008]{Laing2008} 
Laing, R.~A., Bridle, A.~H., Parma, P., \& Murgia, M., 2008, MNRAS, 391, 521

\bibitem[2019]{Loi2019} 
Loi, F., Murgia, M., Govoni, F., et al., 2019, MNRAS, 485, 5285 

\bibitem[2026]{Loi2026} 
Loi, F., Murgia, M., Govoni, F., et al., 2026, A\&A, 710, A96

\bibitem[2007]{Marchegiani2007}
Marchegiani, P., Perola, G. C., \& Colafrancesco, S., 2007, A\&A, 465, 41

\bibitem[2026]{Marchegiani2026}
Marchegiani, P., Murgia, M., Loi, F., et al., 2026, MNRAS, 547, stag352

\bibitem[1998]{Moskalenko98} 
Moskalenko, I.V., \& Strong, A.W., 1998, ApJ, 493, 694

\bibitem[2009]{Murgia2009} 
Murgia, M., Govoni, F., Markevitch, M., et al., 2009, A\&A, 499, 679

\bibitem[2022]{Osinga2022} 
Osinga, E., van Weeren, R.~J., Andrade-Santos, F., et al., 2022, A\&A , 665, A71

\bibitem[2025]{Osinga2025} 
Osinga, E., van Weeren, R.~J., Rudnick, L., et al., 2025, A\&A, 694, A44

\bibitem[2017]{Pinzke2017} 
Pinzke, A., Oh, S.~P., \& Pfrommer, C., 2017, MNRAS, 465, 4800

\bibitem[1999]{Sarazin1999}
Sarazin, C.L., 1999, ApJ, 520, 529

\bibitem[2026]{Sarkar2026} 
Sarkar, A., Miller, E.~D., McNamara, B., et al., 2026, ApJL, 1005, L73

\bibitem[1993]{Tribble1993} 
Tribble, P. C., 1993, MNRAS, 263, 31

\bibitem[2010]{Vacca2010} 
Vacca, V., Murgia, M., Govoni, F., et al., 2010, A\&A, 514, A71

\bibitem[2012]{Vacca2012} 
Vacca, V., Murgia, M., Govoni, F., et al., 2012, A\&A, 540, A38

\bibitem[2026]{Vacca2026} 
Vacca, V., Govoni, F, Murgia, M., et al., 2026, in Advancing Astrophysics with the SKA – II (AASKAII). Report number AASKAII/Vacca01, arXiv:2606.25616

\bibitem[2019]{vanWeeren2019}
van Weeren, R. J., de Gasperin, F., Akamatsu, H., et al., 2019, Space Science Reviews, 215, 16

\bibitem[2026]{vanWeeren2026} 
van Weeren, R.~J., Osinga, E., Brunetti, G., et al., 2026, MNRAS, 546, stag054 (W26)

\bibitem[2012]{Vazza2012} 
Vazza, F., Roediger, E., \& Br{\"u}ggen, M., 2012, A\&A, 544, A103

\bibitem[2006]{Vikhlinin2006} 
Vikhlinin, A., Kravtsov, A., Forman, W., et al., 2006, ApJ, 640, 691

\bibitem[2014]{Zandanel2014}
Zandanel, F., Pfrommer, C., \& Prada, F., 2014, MNRAS, 438, 124

\end{thebibliography}
\end{document}